\documentclass[aps,twocolumn]{revtex4-2}
\usepackage{graphicx}
\usepackage{latexsym}
\usepackage{wasysym}
\usepackage{amssymb}
\usepackage{epstopdf}

\begin{document}

\title{Dynamics of Polymer Ejection from a Nano-sphere}

\author{Farzaneh Moazemi}
\affiliation{Institute of Nanoscience and Nanotechnology, University of Kashan, Kashan 87317-53153, Iran}

\author{Samaneh Ghanbari-Kashan}
\affiliation{Institute of Nanoscience and Nanotechnology, University of Kashan, Kashan 87317-53153, Iran}

\author{Narges Nikoofard}
\email{nikoofard@kashanu.ac.ir}
\affiliation{Institute of Nanoscience and Nanotechnology, University of Kashan, Kashan 87317-53153, Iran}

\date{\today}

\begin{abstract}

Polymer ejection from nano-confinement has been of interest due to its relation to various fundamental sciences and applications. However, the ejection dynamics of a polymer with different persistence lengths from confinement through a nanopore is still poorly understood. In this manuscript, a theory is developed for the ejection dynamics of a polymer with the total length $L_0$ and persistence length $l$ from a sphere of diameter $D$.
% During ejection, the contour length of the polymer inside the sphere evolves with time, and the polymer undergoes between two to three confinement regimes, in some cases. Besides, the free energy of polymer attachment to the sphere surface governs the polymer dynamics, at the final stages of the ejection. 
These length-scales specify different regimes, which determine the polymer dynamics and its ejection rate. It is seen that the polymer undergoes between two to three confinement regimes, in some cases. The total ejection time $\tau$ depends on the polymer dynamics in various relevant regimes that the polymer experiences.
Dependence of the ejection time on the system parameters is discussed according to the theory. The theory predicts that $\alpha$ in $\tau \sim L_0^{\alpha}$ changes between 1 and 1.7, $\beta$ in $\tau \sim D^{\beta}$ changes between 3 and 5, and  $\gamma$ in $\tau \sim l^{\gamma}$ is often smaller than 1, in the studied range of the parameters.

\end{abstract}

\maketitle

\section{Introduction}\label{sec1}

Advances in nano-technology have enabled confinement of polymers in various nano-structures \cite{liu2015,cadinu2017,zhou2017,marie2018}. This has new efforts to improve theories for describing the behavior of confined polymers \cite{nikoofard2015,werner2017,gupta2018,schotzinger2020}. 
On the other hand, polymer confinement is a ubiquitous phenomenon in nature. In a eukaryotic cell, biopolymers are often confined in different organelles, including the nucleus \cite{amitai2017}. The genetic material is also confined in bacterias \cite{pelletier2012} and viruses \cite{molineux2013}. 
Confinement of polymers has also notably important applications, such as long DNA sequencing \cite{zrehen2019}, data storage \cite{cao2020} and polymer separation \cite{Heidari2020}.

Previous studies show that the polymers have different properties when they are confined in nano-scale geometries \cite{bleha2018,morrin2021,teng2021}. The effect of confinement is more complicated when the confined polymer has a persistence length comparable to the size of the confining geometry \cite{sakaue2007}. This is due to the existence of several length scales in the system. Among different natural and synthetic polymers, double-stranded DNA has the largest persistence length. 

Compaction and ejection of a polymer from the closed geometry of a sphere is related to natural phenomena, such as viral genome packaging and ejection \cite{keller2016,li2021}. 
Ejection of a confined polymer from a sphere has been elucidated theoretically, by using computer simulations and experimentally \cite{prl2019,park2021, kellermayer2018}. These studies consider how the system parameters such as the size of the confining geometry, the polymer length, the packaging history or the triggering force affect the ejection process. The effect of persistence length and DNA cholesteric interactions are considered using computer simulations \cite{marenduzzo2013,linna2017}. However, a theory that describes dependence of the ejection process on the persistence length of the polymer is lacking, to the authors' knowledge. 

Initial theories for polymer ejection from confinement assumed that the ejection time is proportional to the inverse of the free energy of the polymer in confinement \cite{luijten2006}. Sakaue and Yoshinaga presented another theory, in which the rate of change of the free energy of confinement is balanced with the rate of energy dissipation in the system \cite{sakaue2009}. Later, this theory was improved by revising the final stage of the polymer ejection, when there is no free energy of confinement \cite{prl2019}.

In this manuscript, the theory is revisited to describe the ejection dynamics of a polymer, which has an arbitrary persistence length. It is shown that considering the persistence length changes the theory of polymer ejection, significantly. 
Predictions of the theory for dependence of the ejection time on the total length of the polymer, size of the confining geometry and the persistence length are discussed.

\begin{figure*}
\begin{center}$
\begin{array}{cc}
\includegraphics[width=8cm,height=6cm]{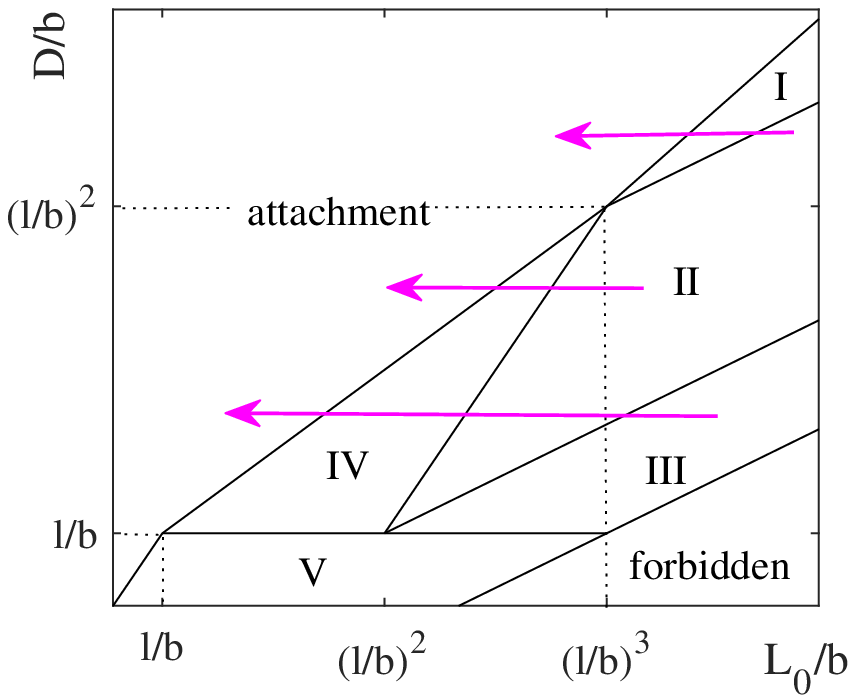}&
\includegraphics[width=10cm,height=8cm]{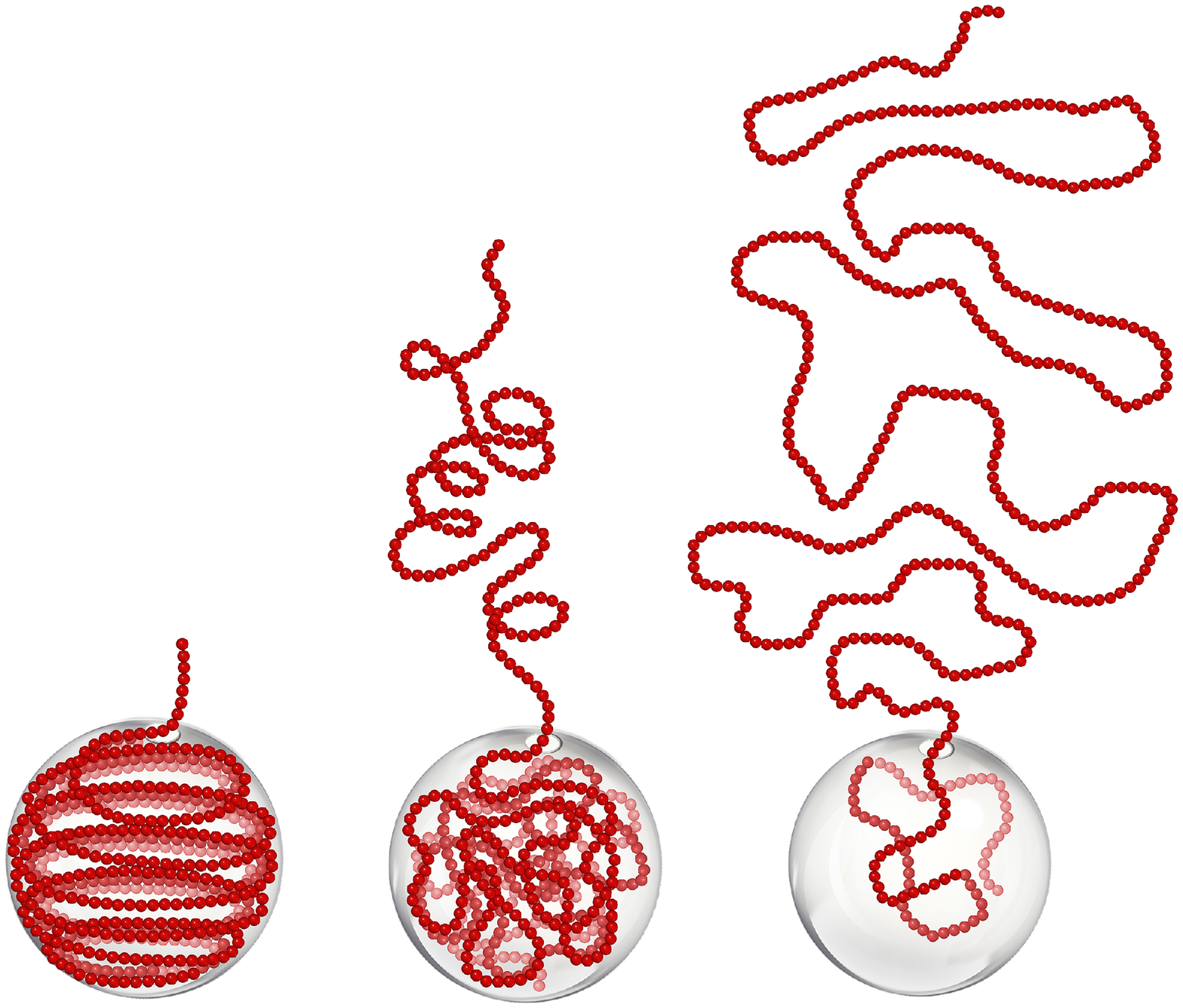}
\end{array}$
\end{center}
\caption{Left: Phase diagram shows the different confinement and the attachment regimes for a polymer of length $L_0$ and Kuhn length $l$ confined in a nano-sphere of size $D$. During ejection, the polymer length inside the sphere decreases and the system moves along the x-axis in the phase diagram. The arrows show some typical paths in which the polymer experiences several regimes, during its ejection. Right: The polymer starts ejection, from the regime III with an ordered state. Then, it experiences the regimes II and afterwards the regime IV, before complete ejection from the sphere. The polymer in the sphere is shown in the regimes III, II and IV, from left to right, respectively.} \label{fig1}
\end{figure*}

\section{Free Energy of a Semi-flexible Polymer in a nano-sphere}\label{sec2}
In this section, the theory of a semi-flexible polymer confined inside a sphere is reviewed briefly, as an integral part of the study. A semi-flexible polymer is effectively described by cylindrical Kuhn monomers. For a polymer with persistence length $P$, the Kuhn monomers have length $l=2P$ and width $b$. Number of the Kuhn monomers is equal to $N=\frac{L}{l}$, where $L$ is contour length of the polymer. Radius of gyration of a free polymer is described by $R_g\approx v^{\frac{1}{5}}l^{\frac{2}{5}}N^{\frac{3}{5}}$; in which $v=l^2b$ is the second virial coefficient \cite{rubinstein2003}. A polymer with persistence length $P$ and contour length $L$ confined in a closed cavity of size $D$ ($<R_g$) is below described in five different regimes (Fig. \ref{fig1}) \cite{sakaue2007}.

%\section{Results}\label{sec2}

%Sample body text. Sample body text. Sample body text. Sample body text. Sample body text. Sample body text. Sample body text. Sample body text.

%\section{This is an example for first level head---section head}\label{sec3}

\subsection{Fluctuating Semi-dilute Regime}\label{regimeI}
\begin{figure*}
 \centering
 \includegraphics[scale=0.5]{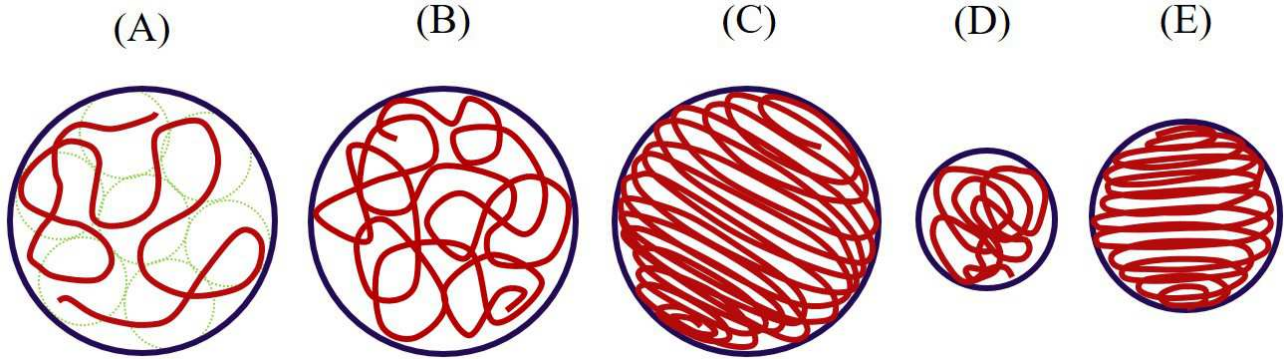}
 \caption{Configurations of a semi-flexible polymer inside a nano-sphere. (a) In the regime I, the polymer is divided into confinement blobs. The polymer length inside each blob does not feel confinement. (b) In the regime II, the polymer density inside the sphere increases. This results in an ideal behavior for the polymer, in this regime. (c) The density further increases in the regime III. Thus, entropic forces between different parts of the polymer cause an ordered state in the polymer configuration. (d) In the regime IV, the radius of the sphere is smaller. The system is no longer uniform, unlike a polymer in the bulk. (e) In the regime V, bending energy governs the polymer behavior and results in another ordered state.
 }
 \label{1TT}
\end{figure*}

In this regime, the polymer is divided into confinement blobs. Inside each blob, statistics of the monomers is not perturbed by the confinement (Fig. \ref{1TT}(a)). Thus, dependence of the size of the blobs $\xi$ on the contour length inside each blob $L_b$ is
\begin{equation} \label{bs}
 \xi \approx l^{\frac{1}{5}}b^{\frac{1}{5}}L_b^{\frac{3}{5}}.
\end{equation} 
The blobs are closely packed inside the sphere; $\frac{L_b}{\xi^3}\approx \frac{L}{D^3}$. The two former equations give
\begin{equation} \label{LbI}
L_b  \approx \frac{D^{\frac{15}{4} }}{ l^{\frac{3}{4} }  b^{\frac{3}{4} } L^{\frac{5}{4} }}.
\end{equation}  
The free energy of confinement is equal to thermal energy times the number of confinement blobs \cite{sakaue2007}; 
\begin{equation} \label{FI}
F_c \approx k_{B}T \frac{L}{L_b} \approx k_{B}T \frac{l^{\frac{3}{4}}  b^{\frac{3}{4}} L^{\frac{9}{4} }}{D^{\frac{15}{4}}}.
\end{equation}

The sphere size should be smaller than the radius of gyration of the polymer in the bulk, $D<R_g$. Besides, the confinement blob should be larger than thermal blob, $\xi>\xi_T$. Thermal blob, $\xi_T$ is a length scale in the system. On smaller length scales, the excluded volume interaction is smaller than the thermal energy. Indeed, the excluded volume interactions between the confinement blobs are assumed in the derivation of Eq. \ref{LbI}. 
In the above conditions, substituting $R_g$, $\xi$ and  $\xi_T\approx \frac{l^2}{b}$ results that the regime I is true within certain boundaries \cite{sakaue2007}
\begin{equation} \label{BI} 
l\left(\frac{L}{b}\right)^{\frac{1}{3}} < D < l^{\frac{1}{5} } b^{\frac{1}{5} } L^{\frac{3}{5} }.
\end{equation}

\subsection{Mean-field Semi-dilute Regime}\label{regimeII}
As the density of the polymer in the sphere increases, the blob size becomes smaller than the size of the thermal blob, $\xi < \xi_T$, and there is a crossover to the regime II. In this regime, the polymer behavior is Gaussian at all length scales (Fig. \ref{1TT}(b)). Besides, the mean-field-approximation is valid and the free energy is calculated from the second term of the virial expansion \cite{sakaue2007}
\begin{equation} \label{FII}
F_c \approx vc^2 \approx k_{B} T \frac{L^2 b}{D^3}. 
\end{equation}
Here, $c=\frac{L}{lD^3}$ is concentration of the Kuhn monomers inside the sphere. The length scale at which the concentration fluctuations can be ignored is called the correlation length, and is obtained from the random phase approximation; $\xi \approx \left(\frac{l^2}{12cv} \right)^{\frac{1}{2}} \approx \left(\frac{D^3l}{Lb} \right)^{\frac{1}{2}}$ \cite{sakaue2007}. 
 The contour length related to the correlation length is obtained using $\xi \approx \left(lL_b\right)^{\frac{1}{2}}$; 
 \begin{equation} \label{LbII}
L_b \approx \frac{D^3}{Lb}. 
 \end{equation}
Each correlation length should contain several monomers and be smaller than the size of the sphere;  $l<\xi <D$. 
These conditions besides the initial equation for this regime, $\xi < \xi_T$,  give the boundaries for the regime II  \cite{sakaue2007}
 \begin{equation} \label{BII} 
\left(Lbl \right)^{\frac{1}{3}}< D< \min\{l \left(\frac{L}{b}\right)^{\frac{1}{3}},\frac{Lb}{l}\}.
 \end{equation}
 
\subsection{Liquid Crystalline Regime}\label{subsec2}
As the density increases, the correlation length becomes comparable to the Kuhn length, $\xi < l$. The system responds by a transition and the coexistence of isotropic and nematic phases (Fig. \ref{1TT}(c)). The free energy of confinement is due to the loss in the orientational entropy, and is estimated by the number of Kuhn segments \cite{sakaue2007}; 
 \begin{equation} \label{FIII}
F_c \approx k_{B} T \frac{L}{l}.
  \end{equation}
In this regime, the Kuhn segments have arranged inside the sphere, and the bending energy of the Kuhn segments is not relevant yet. Thus, the size of the sphere should be larger than the Kuhn length, $D>l$. Obviously, the volume of the sphere should be larger than the total volume of the monomers, $D^3>Lb^2$. Overall, the boundaries for the regime III are \cite{sakaue2007}
 \begin{equation}  \label{BIII} 
\max\{\left(Lb^2\right)^{\frac{1}{3}},l\}<D<\left(Lbl \right)^{\frac{1}{3}}. 
\end{equation}

\subsection{Ideal Chain Regime}\label{subsec2}
When the correlation length of the regime II becomes larger than the size of the system, $\xi\sim \left(\frac{D^3l}{Lb} \right)^{\frac{1}{2}}>D$, the polymer enters the regime IV \cite{sakaue2007}. The correlation length gives a description of  fluctuations in the polymer configuration (Fig. \ref{1TT}(d)). Also, in this regime, the polymer behavior is Gaussian at all length scales. The polymer is divided into blobs, inside which the confinement is not felt by the monomers; $\xi\approx \left(lL_b\right)^{\frac{1}{2}}$. These blobs completely overlap each other and their size is equal to the size of the sphere; $\xi \approx D$. The contour length of the polymer inside the blobs is 
 \begin{equation} \label{LbIV}
L_b \approx \frac{D^2}{l}. 
 \end{equation}
The free energy of confinement results from the loss of the degrees of freedom for each blob; 
 \begin{equation} \label{FIV}
F_c \approx k_BT \frac{L}{L_b} \approx k_BT \frac{lL}{D^2}.
 \end{equation}
In this regime, the sphere size is smaller than the correlation length, however, the bending energy is not determining yet. Thus, the sphere size should be larger than the Kuhn length, $D>l$. Besides, the size of the sphere should also be smaller than the radius of gyration of a free \emph{Gaussian} polymer, in order to have confinement effect. These conditions result in   \cite{sakaue2007}
 \begin{equation} \label{BIV} 
\max\{\frac{Lb}{l},l\}<D<\left(lL\right)^{\frac{1}{2}}.
\end{equation}

\subsection{Bending Regime}\label{subsec2}
When the size of the sphere becomes smaller than the Kuhn length $D<l$, the behavior of the confined polymer is dominated by bending energy  (Fig. \ref{1TT}(e)) \cite{sakaue2007}. In this regime, the Odijk length, $\lambda$ is the contour length of the polymer between two deflections from the sphere surface \cite{odijk1983}; 
 \begin{equation} \label{LbV}
\lambda^3 \approx D^2 l. 
\end{equation}
The free energy is obtained from the bending energy of a polymer of length $L$ and bending modulus $\kappa = k_BTl$ with a radius of curvature of order $D$; 
 \begin{equation} \label{FV}
F_c \approx \kappa \frac{L}{D^2} \approx k_BT \frac{lL}{D^2}.
\end{equation}
This approximation is valid at low volume fractions, $\phi<0.2$. At higher volume fractions, interaction between the Kuhn segments becomes determining. In this regime, the volume of the sphere should be larger than the total volume of the monomers. The boundaries for the regime V are \cite{sakaue2007}
 \begin{equation} \label{BV} 
\left(Lb^2\right)^{\frac{1}{3}}<D<l. 
\end{equation}

%In the regime I, the polymer is distributed in the sphere. In the regime II, the density increases, such that different parts of the polymer are very close to each other. In the regime III, the density is higher and entropic forces arrange the polymer inside the sphere. The regimes IV and V occur in smaller radii of the sphere. In the regime IV, the radius is so small that the polymer is distributed in an asymmetric configuration inside the sphere. In the regime V, the radius becomes smaller than the persistence length and bending energy governs organization of the polymer. 

\begin{figure*}
 \centering
 \includegraphics[scale=0.35]{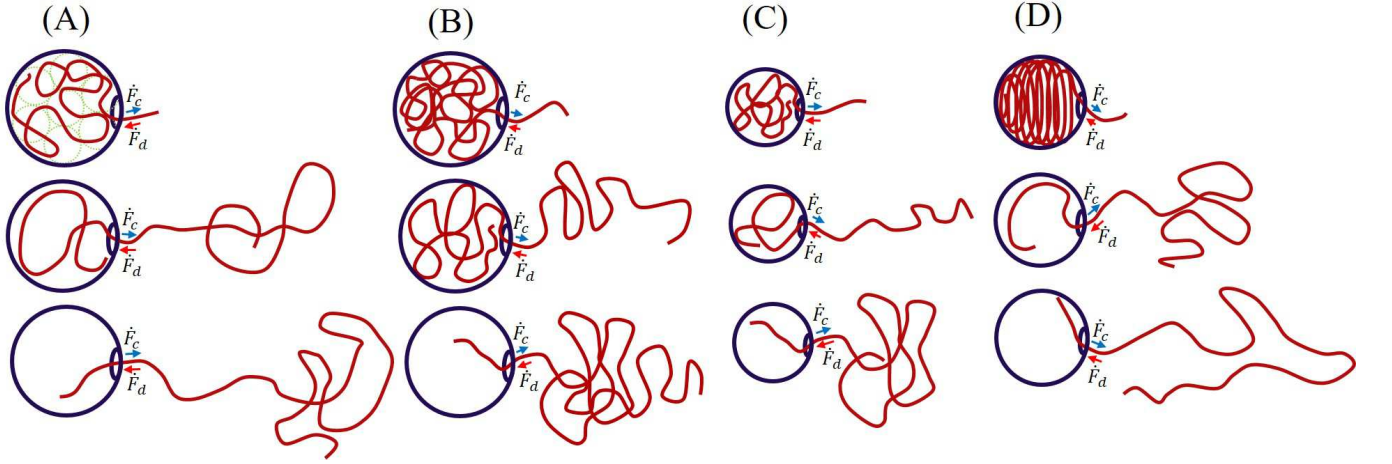}
 \caption{The process of polymer ejection, from the nano-sphere. $\dot{F}_c$ and $\dot{F}_d$ are confinement and drag forces. They are parallel to the polymer contour and are applied on the monomer which is inside the nano-pore. (a) Top: polymer starts ejection, in the regime I. Middle: There has remained one blob inside the sphere. Bottom: The nano-sphere has no confinement effect on the polymer. Attachment to the sphere governs the ejection. (b) Top: polymer starts ejection from regime II. Middle: Polymer experiences the regime I, another confinement regime. Bottom: Attachment of the polymer to the sphere governs the ejection. (c) Top: polymer starts ejection, in the regime IV. Middle: There has remained one blob inside the sphere. The blob is different from the blob in the regime I, since the sphere is smaller. Bottom: Attachment to the sphere governs the ejection. (d) Top: polymer starts ejection, in the regime V. Middle: The polymer that is remained inside the sphere is equal to the radius of the sphere. Bottom: There is no bending energy. Attachment to the sphere governs the ejection.  }
 \label{2TT}
\end{figure*}

\section{Ejection Dynamics of a Semi-flexible Polymer from a nano-sphere}\label{sec3}
In this section, we develop a theory for the ejection dynamics of a semi-flexible polymer from a sphere. 
The contour-length of the polymer remained inside the sphere at time $t$ is shown by $L(t)$. The rate of change of the free energy is balanced by the rate of energy dissipation \cite{sakaue2009},  
\begin{equation} \label{balance}
\dot{F}(t)\approx -\eta l \left[ \dot{L}(t) \right]^2. 
\end{equation}
The right hand side is the rate of energy dissipation; where $\eta$ is the viscosity of the solvent. Similar to previous studies \cite{sakaue2007,prl2019}, it is assumed that one Kuhn length $l$ near the nano-pore contributes to the dissipation. This is because nearly a Kuhn length of the polymer beside the nano-pore is moving and the rest of the polymer does not feel the ejection.

At first, the ejection is driven by the confinement free energy. As the length of the polymer inside the sphere decreases, the free energy of confinement ceases. Then, the free energy of attachment of the polymer to the sphere becomes dominant. The free energy of attachment to a surface results from the reduction in the possible configurations of the chain besides the surface; $k_BT(1-\lambda)\ln(\frac{L}{l})$. Here, $\lambda$ is a constant of order unity and $\frac{L}{l}$ is the number of Kuhn segments in the attached chain. For a polymer escaping through a nano-pore between two compartments, the free energy of attachment has two terms arising from the two sections of the chain on both sides; $F_a(t) \sim k_BT\left[ (1-\lambda_i)\ln(\frac{L(t)}{l}) + (1-\lambda_o)\ln(\frac{L_0-L(t)}{l})\right]$ \cite{prl2019}. 
Using the rate balance (Eq. \ref{balance}) gives $l\dot{L}(t)\sim -\frac{b^3}{\tau_0} \left[\frac{1-\lambda_i}{L(t)}-\frac{1-\lambda_o}{L_0-L(t)} \right]$, where $\tau_0=\frac{\eta b^3}{k_BT}$ and $L_0$ is the total length of the polymer. The second term in the right side is negligible, because $L_0 \gg L(t)$. Solving the resulting differential equation $\dot{L}(t) \sim -\frac{b^3}{\tau_0 l L(t)}$ gives 
\begin{equation} \label{Da}
L(t)\sim L_{\tau} \left(1-\frac{t-\tau}{\tau_a} \right)^{\frac{1}{2}}, \qquad  \qquad {\tau }_{a}\sim \tau_0 \frac{L_{\tau}^2 l}{b^3}. 
\end{equation}
Here, $\tau$ is the crossover time between the confinement and the attachment stages. $L_{\tau}$ is the length of the polymer inside the sphere at time $\tau$ \cite{prl2019}.

In the following, the polymer dynamics is described in the different regimes of confinement. The polymer behavior is simple in the regimes I, IV and V. However, the polymer experiences a combination of different confinement regimes during its ejection, in the regimes II and III. Accordingly, we first discuss the simple regimes, then, the more complicated ones.

\subsection{Regime I}\label{subsec3}
At the beginning, the polymer dynamics is determined by the confinement free energy (Fig. \ref{2TT}(a)). Using the rate balance (Eq. \ref{balance}) and the confinement free energy in the regime I (Eq. \ref{FI}) gives the polymer velocity; $\dot{L}(t)\sim -\frac{1}{\tau_0}l^{-\frac{1}{4}}b^{\frac{15}{4}}D^{-\frac{15}{4}}L(t)^{\frac{5}{4}}$. By integrating this equation, the time evolution of the number of monomers inside the sphere becomes
 \begin{equation} \label{DI}
L(t)\approx L_0\left(1+\frac{t}{{\tau }_{1,I}}\right)^{-4}, \qquad \qquad {\tau }_{1,I}\sim {\tau }_0 \left(\frac{lD^{15}}{b^{15}L_0}\right)^{\frac{1}{4}}.
\end{equation}
This equation is valid until $t<\tau_{2,I}$, that the sphere has no confinement effect on the polymer. At this time, there has remained one blob inside the sphere (Fig. \ref{2TT}(a)). Using Eq. \ref{LbI} to find the contour length inside a blob at this time, one has $L(\tau_{2,I})\approx L_b(\tau_{2,I}) \approx D^{\frac{5}{3}}\left(lb\right)^{-\frac{1}{3}}$. Substituting in Eq. \ref{DI} gives 
$\tau_{2,I}+\tau_{1,I} \sim {\tau }_0 \left(\frac{D^{10}l}{b^{11}}\right)^{\frac{1}{3}}$. 

The polymer continues its ejection, while it is attached to the sphere (Fig. \ref{2TT}(a)). The polymer dynamics at this stage is obtained by substituting $L_{\tau} = L(\tau_{2,I})$ in Eq. \ref{Da};
$L(t)\approx D^{\frac{5}{3}}\left(lb\right)^{-\frac{1}{3}}\left(1-\frac{t-\tau_{2,I}}{\tau_{a,I}}\right)^{\frac{1}{2}}$, where ${\tau }_{a,I}\sim \tau_0 \left(\frac{D^{10}l}{b^{11}}\right)^{\frac{1}{3} }$. 
The  polymer completely leaves the sphere, $L(\tau_I)=0$, at the time 
$\tau_I \approx \tau_{a,I}+\tau_{2,I} \approx {\tau }_0\left[2\left(\frac{lD^{10}}{b^{11}}\right)^{\frac{1}{3}}-\left(\frac{lD^{15}}{b^{15}L_0}\right)^{\frac{1}{4}}\right]$.
The obtained equations  are in relative agreement with Ref. \cite{prl2019}. There are subtle differences, because the dissipation rate in that study depends on the total length of the polymer.

\subsection{Regime IV}\label{subsec3}
At the beginning of the ejection, the confinement free energy determines the polymer dynamics (Fig. \ref{2TT}(c)). Balancing the energy (Eq. \ref{FIV}) and the dissipation terms gives the polymer velocity; $\dot{L}(t)\approx -\frac{1}{\tau_0}\frac{b^3}{D^2}$. Accordingly, one obtains the instant length of the polymer 
\begin{equation} \label{DIV}
L(t)\approx L_0\left(1-\frac{t}{\tau_{1,IV}}\right), \qquad \qquad \tau_{1,IV}\approx \tau_0\frac{L_0D^2}{b^3}. 
\end{equation}
Again, this equation is valid until $t<\tau_{2,IV}$, when the polymer length inside the sphere becomes so small that it does not feel the confinement (Fig. \ref{2TT}(c)). At this time, there has remained one blob inside the sphere; $L(\tau_{2,IV})\approx L_b \approx \frac{D^2}{l}$. Substituting this condition in Eq. \ref{DIV} gives 
$\tau_{1,IV}-\tau_{2,IV}\approx \tau_0\frac{D^4}{lb^3}$. 

The rest of the process continues under the attachment free energy (Fig. \ref{2TT}(c)). Substituting $L(\tau_{2,IV}) \approx \frac{D^2}{l}$ in Eq. \ref{Da} gives the polymer dynamics;
$L(t)\approx \frac{D^2}{l}\left(1-\frac{t-\tau_{2,IV}}{\tau_{a,IV}}\right)^{\frac{1}{2}}$, where ${\tau }_{a,IV}\sim \tau_0 \frac{D^4}{b^3l}$. 
The total ejection time of the polymer from the sphere is the time at which the polymer length becomes zero, in the above equation; $L(\tau_{IV})=0$. The result is
$\tau_{IV} \approx  \tau_{a,IV}+\tau_{2,IV} \approx \tau_0 \left(\frac{D^2L_0}{b^3}\right)$.
It should be noted that the dynamics is different in the attachment stages of the regimes I and IV. This is because the polymer lengths at the crossover times from the confinement to the attachment stage, $L(\tau_{2,I})$ and $L(\tau_{2,IV})$, are different.

\subsection{Regime V}\label{subsec3}
Substituting the derivative of the confinement free energy (Eq. \ref{FV}) in the rate balance (Eg. \ref{balance})  gives the velocity of polymer ejection from the sphere; $\dot{L}(t)\approx -\frac{1}{\tau_0} \frac{b^3}{D^2}$.
Thus, the polymer dynamics is described by
\begin{equation} \label{DV}
L(t)=L_0\left(1-\frac{t}{\tau_{1,V}}\right), \qquad \qquad \tau_{1,V}\approx \tau_0 \frac{L_0D^2}{b^3}.
\end{equation}
Indeed, the ejection is dominated by the bending energy of the polymer from the sphere boundaries. Thus, the equation is valid until $t<\tau_{2,V}$, when the polymer length inside the sphere becomes equal to the sphere diameter (Fig. \ref{2TT}(d)). Substituting this condition, $L(\tau_{2,V})\approx D $, into Eq. \ref{DV} gives
$\tau_{1,V}-\tau_{2,V}\approx \tau_0 \frac{D^3}{b^3}$.
The above equations for the regime V are only applicable in small volume fractions. 

The rest of the polymer ejection continues in the attachment regime (Fig. \ref{2TT}(d)). Using $L(\tau_{2,V})\sim D$ in Eq. \ref{Da} gives the polymer dynamics;
$L(t)\approx D\left(1-\frac{t-\tau_{2,V}}{\tau_{a,V}}\right)^{\frac{1}{2}}$, where ${\tau }_{a,V}\sim \tau_0 \frac{D^2l}{b^3}$. 
The total ejection time from the sphere is obtained using $L(\tau_{V})=0$ in the above equation;
$\tau_V  \approx \tau_{2,V}+\tau_{a,V} \approx \tau_0 \left( \frac{D^2l}{b^3}+\frac{D^2L_0}{b^3}-\frac{D^3}{b^3}\right)$.

\begin{figure*}
 \centering
 \includegraphics[scale=0.65]{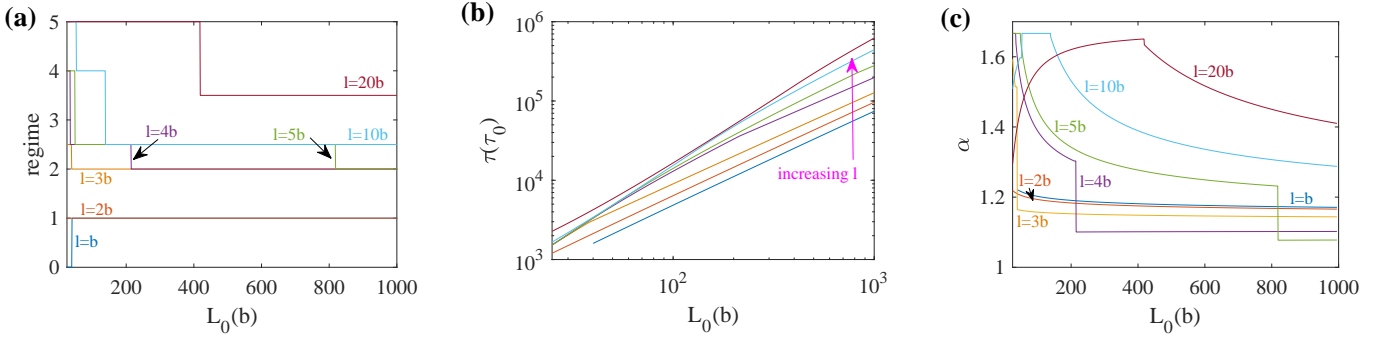}
 \caption{Predictions of the theory for the ejection process versus the polymer length, for different persistence lengths. The volume fraction of the polymer inside the nano-sphere is fixed, $\phi=0.1$. (a) The initial regime of the polymer. (b) Log-log plot of the ejection time versus the polymer length. A linear behavior is observed. (c) The exponent $\alpha$ in the relation $\tau \approx L_0^{\alpha}$, which describes dependence of the ejection time on the persistence length. It is seen that the exponent changes between 1 and 1.7, for all parameters.}
 \label{tN}
\end{figure*}

\subsection{Regime II}\label{subsec3}
When the ejection starts, the polymer dynamics is governed by the confinement free energy of the regime II; Balancing the rate of change of the free energy (the time derivative of Eq. \ref{FII}) with the dissipation rate gives $\dot{L}(t)\sim-{\frac{1}{\tau_0}}{\frac{b^4}{lD^3}}L(t)$.
Solving this differential equation results in the polymer dynamics; 
\begin{equation} \label{DII} 
L(t) \sim L_0\exp (-\frac{t}{\tau_{1,II}}), \qquad \qquad  \tau_{1,II} \sim \tau_0{\frac{lD^3}{b^4}}. 
\end{equation}
This equation is valid until $t<\tau_{2,II}$, when the density of the polymer in the sphere decreases so much that the polymer enters one of the regimes I and IV.  In other word, one of the upper limits for the regime II (Eq. \ref{BII}) is violated, at this time. %, depending on the ratio $\frac{Db}{l^2}$. 

\textbf{Case 1, $\frac{Db}{l^2}>1$:} When $L(\tau_{2,II})\sim b \left(\frac{D}{l}\right)^3$, the driven process in the regime II ceases and the system enters the regime I (Fig. \ref{2TT}(b)). Substituting this condition into Eq. \ref{DII}, one has
$\tau_{2,II} \sim \tau_0{\frac{lD^3}{b^4}} \log ({\frac{L_0}{b}}(\frac{l}{D})^3 )$.  
The polymer dynamics in the regime I is obtained by substituting $L(\tau_{2,II})$ for $L_0$ in Eqs. \ref{DI};
$L(t)\sim b(\frac{D}{l})^3 (1+\frac{t-\tau_{2,II}}{\tau_{1,I}})^{-4}$, where $\tau_{1,I}\sim \tau_0\frac{lD^3}{b^4}$. 
The relation is valid in the time interval $\tau_{2,II}<t<\tau_{3,II}$, in which $\tau_{3,II}$ is the time that the polymer ejection is no longer driven by the confinement energy. The required condition is that there has remained one blob inside the sphere, $L(\tau_{3,II})\approx D^{\frac{5}{3}}\left(lb\right)^{-\frac{1}{3}}$. This relation with the latter equation for polymer dynamics in the regime I gives 
$\tau_{1,I}+\tau_{3,II}-\tau_{2,II}\sim \tau_0(\frac{lD^{10}}{b^{11}})^{\frac{1}{3}}$ . 

The ejection continues under the dominance of the attachment free energy, similar to the regime I (Fig. \ref{2TT}(b)).
Thus, the polymer dynamics is $L(t) \sim \frac{D^{\frac{5}{3}}}{(lb)^{\frac{1}{3}}}\left(1-\frac{t-\tau_{3,II}}{\tau_{a,I}}\right)^{\frac{1}{2}}$, in which $\tau_{a,I}\sim \tau_0\left(\frac{D^{10}l}{b^{11}}\right)^{\frac{1}{3}}$.
Using $L(\tau_{II})=0$ in this equation, the total ejection time from the sphere becomes
$\tau_{II} \sim \tau_{3,II}+\tau_{a,I} \sim  2\tau_0(\frac{lD^{10}}{b^{11}})^{\frac{1}{3}}+\tau_0{\frac{lD^3}{b^4}}{\log({\frac{L_0}{b}}(\frac{l}{D})^3)}-\tau_0{\frac{lD^3}{b^4}}$.

 \textbf{Case 2, $\frac{Db}{l^2}<1$:} 
When $L(\tau_{2,II}){\frac{b}{l}} \sim D$, the driven process in the regime II ceases and the system enters the regime IV. Substituting this condition into Eq. \ref{DII} gives the crossover time $\tau_{2,II}\sim  \tau_0{\frac{lD^3}{b^4}}{\log(\frac{L_0b}{Dl})}$.
Using $L(\tau_{2,II})$ instead of $L_0$ in Eqs. \ref{DIV} gives the polymer dynamics in the regime IV;
$L(t)\sim {\frac{Dl}{b}}(1-{\frac{t-\tau_{2,II}}{\tau_{1,IV}}})$, with $\tau_{1,IV}\sim \tau_0{\frac{lD^3}{b^4}}$. 
This relation is valid in the time interval $\tau_{2,II}<t<\tau_{3,II}$. At time $\tau_{3,II}$, the driven process ceases, because there has remained one blob inside the sphere; $L(\tau_{3,II})\approx \frac{D^2}{l}$. Using this condition in the latter equation for polymer dynamics gives $\tau_{1,IV}-\tau_{3,II}+\tau_{2,II}\sim \tau_0{\frac{D^4}{lb^3}}$. 
The rest of the ejection is dominated by the attachment free energy, and the dynamics is similar to that of the regime IV. After minor calculations, the total ejection time from the sphere becomes $\tau_{II} \sim \tau_{3,II}+\tau_{a,IV} \sim {\tau_0\frac{D^3l}{b^4}}\left(1+\log\frac{L_0b}{Dl}\right)$.

\subsection{Regime III}\label{subsec3}
The rate of change of the free energy is obtained by taking the derivative of Eq. \ref{FIII}. Balancing the energy and the dissipation rates gives the polymer velocity $\dot{L}(t)\sim -\frac{1}{\tau_0}\frac{b^3}{l^2}$. From this relation, the number of remaining monomers inside the sphere at time $t$ is 
 \begin{equation} \label{DIII}
L(t)\approx L_0\left(1-\frac{t}{\tau_{1,III}}\right), \qquad \qquad \tau_{1,III}\approx \tau_0 \frac{l^2L_0}{b^3}. 
\end{equation}
This relation is valid until $t<\tau_{2,III}$, when the polymer density inside the sphere decreases so much that the system experiences the regime II. Indeed, as the contour length of the polymer inside the sphere decreases, the polymer first enters the regime II and then one of the regimes I or IV (Fig. \ref{fig1}). 

Entering the regime II occurs when the upper boundary for $D$ (Eq. \ref{BIII}) is violated; $L(\tau_{2,III})\sim\frac{D^3}{lb}$. This condition with Eqs. \ref{DIII} gives
$\tau_{1,III}-\tau_{2,III}\approx \tau_0\frac{D^3l}{b^4}$. 
Substituting $L(\tau_{2,III})$ for $L_0$ into Eqs. \ref{DII} for the regime II gives
$L(t)\sim \frac{D^3}{lb} \exp(-\frac{{t-\tau_{2,III}}}{\tau_{1,II}})$, in which $\tau_{1,II}\sim \tau_0{\frac{l{D^3}}{b^4}}$. 
This relation is valid in the time interval $\tau_{2,III}<t<\tau_{3,III}$, when the  polymer enters one of the regimes I or IV.

\textbf{Case 1, $\frac{Db}{l^2}>1$:}  
The polymer enters the regime I and then experiences the attachment stage, until it completely leaves the sphere. The polymer dynamics and the crossover times at these two final stages are similar to those of the case 1 of the regime II. 
The total ejection time becomes
$\tau_{III}\sim  {2\tau_0(\frac{l{D^{10}}}{b^{11}})^{\frac{1}{3}}}+2\tau_0{\frac{lD^3}{b^4}}\left(-1+\log{\frac{l}{b}}\right)+\tau_0{\frac{{L_0}{l^2}}{b^3}}$.

\textbf{Case 2, $\frac{Db}{l^2}<1$:} 
The polymer enters the regime IV and then experiences the attachment stage, until its ejection ends. The polymer dynamics and the crossover times in the regime IV and the attachment stage are similar to those of the case 2 of the regime II.
The total ejection time becomes
$\tau_{III} \sim 2\tau_0{\frac{l{D^3}}{b^4}}\log{\frac{D}{l}}+\tau_0{\frac{L_0{l^2}}{b^3}}$.

%\subsubsection{This is an example for third level head---subsubsection head}\label{subsubsec2}

%Sample body text. Sample body text. Sample body text. Sample body text. Sample body text. Sample body text. Sample body text. Sample body text.

\section{Predictions of the theory}\label{sec4}
In the previous section, the theory was obtained by using scaling calculations. Besides, the ejection time for each regime contains several terms, so, it requires many fitting parameters. To the authors' knowledge, the effect of persistence length on the ejection time of a polymer confined inside a sphere has been investigated in one previous study, by using computer simulations \cite{linna2017}. Unfortunately, the simulation data is not enough to find all the fitting parameters, to validate the theory. However, predictions of the theory are given here, for comparison with future simulations and experiments.
In the following, the volume fraction of the polymer inside the sphere $\phi = \frac{L_0b^2}{D^3}=0.1$ is kept fixed, according to the related simulations \cite{linna2017}. 

\begin{figure}
 \centering
 \includegraphics[scale=0.6]{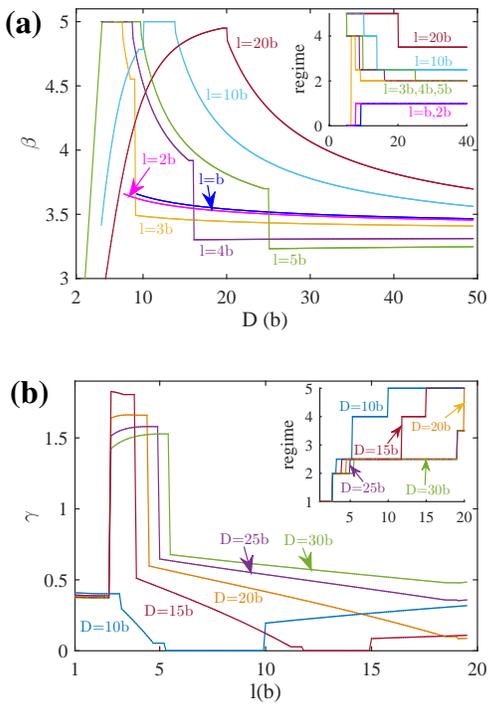}
 \caption{Dependence of the ejection time on the sphere diameter and persistence length, for volume fraction $\phi=0.1$. (a) The exponent $\beta$ that describes dependence of the ejection time on the sphere diameter. It is seen that $\beta$ takes values between 3 and 5, for different sphere diameters and persistence lengths. Inset: the relevant regime for the studied parameters. (b) The exponent $\gamma$ that describes dependence of the ejection time on the persistence length. It is seen that $\gamma$ takes values between 0 and 1, for most of the sphere diameters and the persistence lengths. Inset: the relevant regime for the studied parameters.}
 \label{tDP}
\end{figure}

The predictions of the theory for dependence of the ejection process on the system parameters are summarized in Figs. \ref{tN} and \ref{tDP}. 
Persistence length of the double-stranded DNA is equal to 50nm, while its width is equal to 2.5nm. Thus, we can take $b=2.5nm$ and $l=20b$. Other synthetic and natural polymers have smaller persistence lengths \cite{nikoofard2011}. As a result, we study persistence lengths below 20b, in the following. The largest length of the polymer in this study is $L_0=1000b$. Larger polymers are also tested and the same behavior is observed.

Figure \ref{tN}(a) shows initial regime versus the polymer length, for different persistence lengths. 
On the plots, the values for cases 1 and 2 of the regime II (III) are shown by 2 (3) and 2.5 (3.5), respectively. The zero value for the regime describes the condition in which the sphere has no confining effect on the polymer. It should be noted that the diameter of the sphere changes with the total length of the polymer, to keep the volume fraction fixed. Thus, one cannot deduce the regimes that the polymer experiences during its ejection, from Fig. \ref{tN}(a). It is seen that both persistence and polymer lengths determine the initial regime of the polymer. The initial regime determines different aspects of the statics and the dynamics of the semi-flexible polymer.

Figure \ref{tN}(b) represents the log-log plot of the ejection time versus the polymer length. The ejection time is an increasing function of the persistence and the polymer lengths. Dependence of the ejection time on the system parameters is simple in e.g. the regime IV, by using the related equation. However, dependence of the the ejection time on the system parameters is not easily deduced from the equations of Sec. \ref{sec3}, in the other regimes. Instead, it is possible to use local slope of the curves for the ejection time, in Fig. \ref{tN}(b). 

Figure \ref{tN}(c) shows the exponent $\alpha$ in the relation $\tau \approx L_0^{\alpha}$. It is calculated by using the local slope of the curves in Fig. \ref{tN}(b). The value of $\alpha$ changes between 1 and 1.7 for all studied parameters. It is seen that $\alpha$ is a decreasing function of the polymer length in the regime I, the regime II case 2 and the regime III case 2. It is constant with respect to the polymer length in the regime II case 1 and the regime IV. $\alpha$ increases with the polymer length in the regime V.

Figure \ref{tDP}(a) shows the exponent $\beta$ that describes dependence of the ejection time on the sphere diameter, $\tau \approx D^{\beta}$. The exponent is calculated for various persistence lengths and sphere diameters, in which the polymer experiences different initial regimes (inset of Fig. \ref{tDP}(a)). It is observed that the exponent $\beta$ takes values between 3 and 5. The range of values for $\beta$ is different from those of $\alpha$. Thus, dependence of the ejection time on the sphere diameter is stronger than its dependence on the polymer length.

Besides, Fig. \ref{tDP}(a) shows that $\beta$ is a decreasing function of the sphere diameter in the regime I, the regime II case 2 and the regime III case 2. It is constant with respect to the sphere diameter in the regime II case 1 and the regime IV. $\beta$ increases with the sphere diameter in the regime V. It should be noted that the polymer length and the sphere diameter depend on each other, through the relation $\phi = \frac{L_0b^2}{D^3}$. Thus a similar behavior is observed for $\alpha$ and $\beta$.

Figure \ref{tDP}(b) shows the exponent $\gamma$ that describes dependence on the persistence length, $\tau \approx l^{\gamma}$. The relevant initial regimes are shown in the inset, for different sphere diameters. It is seen that the polymer falls in the regime II case 1, in a small interval of persistence lengths. Outside this interval, the exponent $\gamma$ takes values smaller than 1. So, dependence of the ejection time on the persistence length is generally weaker than its dependence on the polymer length and the sphere diameter. 

Figure  \ref{tDP}(b)   shows that the ejection time does not depend on the persistence length, in the regime IV. It is also seen that the exponent $\gamma$ is constant in the regime II case 1. It decreases with the persistence length in the regime II case 2. An increasing dependence of the persistence length is observed, in the regime V.

\section{Conclusions}\label{sec5}
Here, the ejection dynamics of a polymer of finite persistence length from a nano-sphere was studied, theoretically. The dynamics and the ejection time were obtained for different parameters of the system. 
In this study, it was assumed that one Kuhn length beside the nano-pore contributes in the dissipation. The authors have also investigated the case that one correlation length in each regime contributes to the dissipation. Assuming one correlation length for the dissipation, it is observed that the ejection time is not an increasing function of the persistence length, which is not in agreement with previous simulations \cite{linna2017}. 

The theory was obtained in the situation that there are no hydrodynamic interactions between the monomers, in accordance with the simulations in Refs. \cite{linna2017,prl2019}. However, it is straightforward to include hydrodynamics in the present formalism. In this situation, the dissipation would occur in the range of a correlation length, not a Kuhn length. Thus, the rate of dissipation would be $T\dot{S}(t) \approx \eta \xi(t) \left[ \dot{L}(t) \right]^2$ \cite{sakaue2007}. Definition of the correlation length is easy in the regimes I, II and IV, which is equal to the blob size. However, the correlation length in the regimes III and V needs more consideration. As an estimate, it is possible to define the correlation length equal to the persistence length and the Odijk length in the regimes III and V, respectively. 

Further simulations are needed to check the formalism, more exactly. On the other hand, it would be interesting to extend the present formalism to study the packaging dynamics of a polymer with finite persistence length into a cavity. This problem is already studied by using computer simulations \cite{polson2019}.

\end{document}